# Large Binocular Telescope Adaptive Optics System: New achievements and perspectives in adaptive optics

S. Esposito*[a], A. Riccardi[a], E. Pinna[a], A. Puglisi[a], F. Quirós-Pacheco[a], C. Arcidiacono[a], M. Xompero[a], R. Briguglio[a], G. Agapito[a], L. Busoni[a], L. Fini[a], J. Argomedo[a], A. Gherardi[a], G. Brusa[b], D. Miller[b], J. C. Guerra[b], P. Stefanini[a], P. Salinari[a]

[a]INAF - Osservatorio Astrofisico di Arcetri, Largo E. Fermi 5, 50125 Firenze, Italy;
[b]LBT Observatory, Univ. of Arizona, 933 North Cherry Ave., Tucson AZ 85721,  USA

## ABSTRACT

The Large Binocular Telescope (LBT) is a unique telescope featuring two co-mounted optical trains with 8.4m primary mirrors. The telescope Adaptive Optics (AO) system uses two innovative key components, namely an adaptive secondary mirror with 672 actuators and a high-order pyramid wave-front sensor. During the on-sky commissioning such a system reached performances never achieved before on large ground-based optical telescopes. Images with 40mas resolution and Strehl Ratios higher than 80% have been acquired in H band (1.6 μm). Such images showed a contrast as high as $10^{-4}$. Based on these results, we compare the performances offered by a Natural Guide Star (NGS) system upgraded with the state-of-the-art technology and those delivered by existing Laser Guide Star (LGS) systems. The comparison, in terms of sky coverage and performances, suggests rethinking the current role ascribed to NGS and LGS in the next generation of AO systems for the 8-10 meter class telescopes and Extremely Large Telescopes (ELTs).

**Keywords:** Large Binocular Telescope, high-order adaptive optics, pyramid sensor, adaptive secondary mirror.

## 1. INTRODUCTION

The purpose of Adaptive Optics (AO) is to correct wave-front aberrations of the electromagnetic waves received from a ground-based telescope after propagation through the atmosphere obtaining an image quality reached otherwise only with space-based telescopes. The Large Binocular Telescope[1] (LBT) features two 8.4m optical trains, each one of them fully integrated with AO correction provided by the two Adaptive Secondary Mirror (ASM) units with 672 actuators. The First-Light AO (FLAO) system[2], designed for each optical train, is a Natural Guide Star (NGS) AO system featuring a modulated pyramid Wave-Front Sensor (WFS) with a maximum pupil sampling of 30×30 subapertures. A so fine spatial sampling is new for astronomical AO typically working in a photon starving regime. The spatial sampling is difficult to be realized also because of the technical limitations imposed by the limited degrees of freedom available on wave-front correctors. In the compromise between correction performance and available light, the efficiency of wave-front sensors plays a dominant role. In our system, the use of state-of-the-art devices such as the pyramid WFS and the adaptive secondary mirror offers the optimal playground for very high Strehl Ratio (SR) corrections on the bright end, and high efficiency use of reference starlight photons on the faint one. As we will show in this paper, we have obtained >80% correction @ 1.6 μm on a bright star (up to $M_R$~9), and closed efficiently the loop on a ~17[th] magnitude star with ~15 photons per subaperture per frame (~500 in the whole pupil).

The on-sky commissioning of the FLAO#1 system (installed on the right side of the telescope) started in May 2010 and will be completed in the fall of 2011. The installation of the FLAO#2 system on the left side of the telescope started in August 2011. The two FLAO systems will provide AO correction on the bent-Gregorian focal stations serving the LUCIFER spectro-imagers[4].

We will present in this paper the on-sky performance analysis of the FLAO#1 system based on the data collected during the commissioning nights. After a brief system overview, we will present and discuss in the paper the performance achieved during the commissioning campaign. Finally we will present some considerations for future AO systems based on the results achieved at LBT.

*esposito@arcetri.astro.it; phone +39 055 2752 309; fax +39 055 2752 292

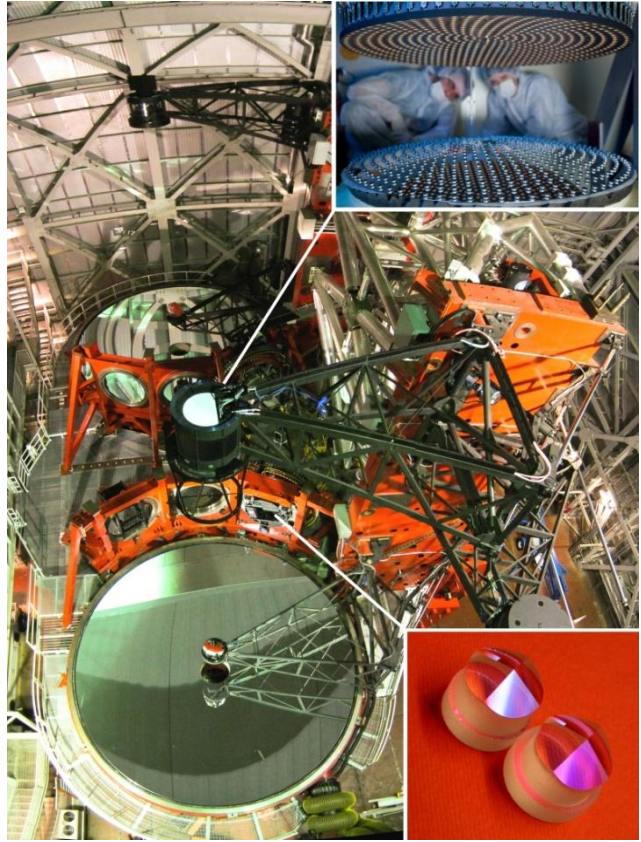

Figure 1. FLAO#1 system installed on the right-side of the Large Binocular Telescope. The Acquisition, Guiding and Wavefront sensing (AGW) unit containing the pyramid WFS is located on the front-right bent-Gregorian focal station. In this photograph the ASM unit features the retro-reflector optics used for system calibrations mounted below it.

## 2. FLAO SYSTEM OVERVIEW

We will briefly review in this section the general characteristics of the FLAO#1 system as it was commissioned to the telescope (Figure 1). A detailed description of the system optical design and laboratory characterization can be found in the cited papers[2][3].

The ASM is equipped with a Zerodur thin shell of 1.6 mm thickness and 911 mm diameter. It relies on 672 contactless voice-coil force actuators to shape the thin shell with a maximum stroke of ~100μm. The settling time ranges from 0.7 to 1ms, and the best figuring error obtained at the telescope is ~28 nm RMS in wavefront, using approximately the 25% of the forces range[5]. The FLAO#1 commissioning results have been obtained with a total of 641 actuators. The contactless technology allows to software deactivate the 31 defective actuators without introducing hard points on the shell.

The pyramid WFS[6] features a maximum of 30×30 subapertures to sample the mirror shape. The equivalent sampling on the primary mirror is 0.28m which matches the Fried parameter ($r_0$) obtained under good seeing conditions at the WFS central wavelength of $\lambda_s$ = 750nm. The WFS camera uses a fast readout E2V CCD39 unit with 80x80 pixels and a maximum frame rate at full frame of $f_s$ =1 kHz.

The temporal and spatial sampling of the WFS can be adjusted as a function of $M_R$, the equivalent R-magnitude of the guide star (§3.1). In particular, the number of subapertures across the pupil diameter can be easily changed by choosing one of the on-chip CCD binning modes. In this way, the telescope pupil can be sampled with 30×30, 15×15, 10×10, or ~7×7 subapertures when using, respectively, binning modes #1, #2, #3, or #4. The sensitivity of the WFS can also be adjusted by means of the pyramid (circular) modulation.

Table 1. Typical system configurations used on sky as a function of the equivalent R-magnitude of the GS ($M_R$). The system parameters are: the binning mode, the temporal sampling frequency ($f_s$), the number of controlled modes ($n_{mod}$), and the pyramid modulation. Median RON values measured for each configuration are also listed.

| $M_R$ | Binning mode | *Pupil sampling (# subaps.)* | $f_s$ (Hz) | $n_{mod}$ | Pyr. mod. ($\pm\lambda_s/D$) | RON ($\sigma_{e^-}$) |
|---|---|---|---|---|---|---|
| $M_R \leq 8.0$ | 1 | 30×30 | 990 | 500 | 2.0 | 10.5 |
| $M_R \leq 10.0$ | 1 | 30×30 | 990 | 400 | 3.0 | 10.5 |
| $10.0 < M_R < 13.5$ | 2 | 15×15 | $990 \leq f_s \leq 300$ | 153 | 3.0 | 6.4 |
| $13.5 \leq M_R < 14.5$ | 3 | 10×10 | $500 \leq f_s \leq 200$ | 66 | 6.0 | 4.5 |
| $14.5 \leq M_R < 16.5$ | 4 | ~7×7 | $400 \leq f_s \leq 100$ | 36 | 6.0 | 4.6 |
| $16.5 \leq M_R < 18.0$ | 4 | ~7×7 | 100 | 10 | 6.0 | 4.6 |

The typical system parameters used on sky as a function of $M_R$ are summarized in Table 1. Also in this table are listed the actual Read-Out Noise (RON) median values measured for each system configuration.

The FLAO control system implementation is based on a modal approach. To be more effective in correcting the turbulence, the modal control basis uses Karhunen-Loève (KL) modes fitted by the ASM[7]. The temporal controller is a modal integrator. Since the optimal integrator gains strongly depend on the atmospheric conditions, a procedure to automatically estimate the optimal gains has been developed. This procedure is run just after the Guide Star (GS) is acquired. Currently, gain values are not further changed during normal AO operation.

The modal interaction matrices between the ASM and the WFS (one for each pair of binning mode and pyramid modulation values) were calibrated at the telescope on daytime with the use of the retro-reflector optics[2] installed below the ASM permitting the calibration beam to reach back the WFS unit (Figure 1).

## 3. OBSERVING CONDITIONS

The maximum performance that can be attained with an AO system is limited by several factors, the most important being the total received flux from the GS, the atmospheric turbulence conditions (i.e. the seeing value, and the wind speed), the separation between the object of interest and the GS, the read-out noise of the WFS detector, and the telescope vibrations. We will focus in this paper on the on-axis performance results of the FLAO#1 system in terms of the SR in H band as a function of the star magnitude and the atmospheric seeing value. We will describe in this section how we have estimated these two important parameters.

### 3.1 Equivalent star magnitude in R band

The FLAO's pyramid WFS is sensitive to a broadband encompassing R and I bands. On the lower bound, the passband is limited by a dichroic sending all the light above 600nm to the WFS. On the upper bound, the quantum efficiency curve of the WFS detector —a deep-depletion CCD39— limits the passband to ~950nm. Therefore, the luminosity of the GS in this band, 600–950nm, drives the AO correction capabilities.

For the purpose of comparing the performance of the FLAO system obtained with different reference stars, we will use the concept of an "equivalent" star magnitude in R band computed as:

$$M_R = -2.5 \log_{10}\left[\frac{h \cdot c}{\tau_{sys} \cdot \lambda_s \cdot e_0} n_{ph}\right] \quad (1)$$

where $n_{ph}$ stands for the number of photons per squared-meter per second [ph m$^{-2}$ s$^{-1}$]. This quantity can be easily estimated from the acquired WFS CCD39 frames. The constants in Eq. (1) are: $h$, the Planck constant; $c$, the light velocity; $e_0 = 1.76 \cdot 10^{-8}$ W m$^{-2}$ µm$^{-1}$, the zero-magnitude brightness in the Johnson R band; $\lambda_s = 750$nm, the WFS central wavelength; and $\tau_{sys}$, the overall optical transmission (Telescope + WFS board + CCD39) in the spectral band 600–950nm.

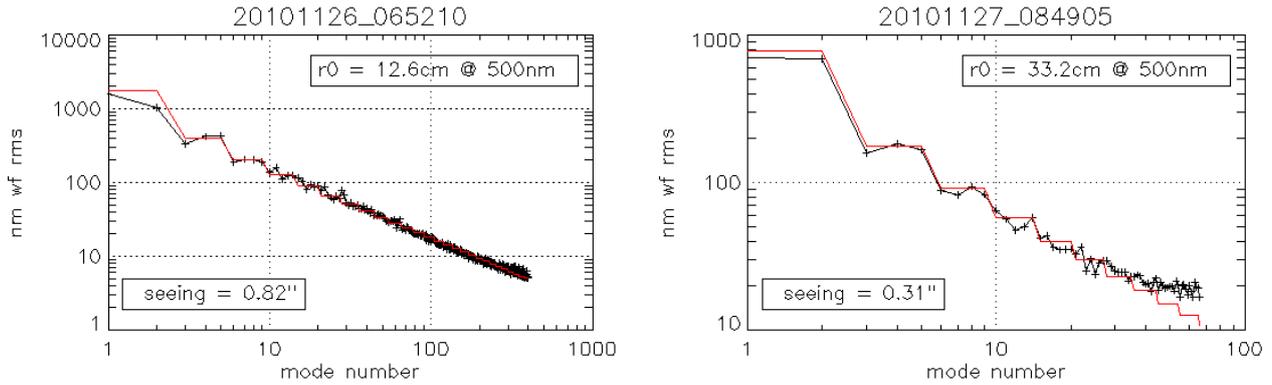

Figure 2. Example of $r_0$ fitting from reconstructed OL modes. (Crosses) Estimated OL modal variances; (Red line) Theoretical modal variance distribution. (*Left*) Case of bin#1, 400 controlled modes. (*Right*) Case of bin#3, 66 controlled modes.

**3.2 Estimation of the seeing value**

In order to have an estimate of the turbulence conditions during a given observation, LBT uses a Differential Image Motion Monitor (DIMM) installed on top of the telescope mount, and an anemometer installed on top of the telescope enclosure. An additional estimate of the seeing value can be found using the AO system telemetry data. This on-line seeing estimation method is based on the reconstruction of Open-Loop (OL) modes (i.e. the modal decomposition of the turbulence) from applied ASM commands and residual WFS signals acquired during the closed-loop operation of the AO system[8]. Once the OL modes are available, the Fried parameter ($r_0$) can be estimated from the OL modal variance distribution, from which the seeing value can be simply computed as $s(\lambda)=0.98\lambda/r_0$ (by convention $s(\lambda)$ quoted at $\lambda=500$nm).

Figure 2 shows the seeing estimation results for two particular cases. The plots show the time-variance distribution of the reconstructed OL modes, and the best-fit theoretical (Kolmogorov) variance distribution of Zernike modes. In order to avoid the uncertainty of the outer scale of the turbulence ($L_0$) and the residual telescope vibrations, the first three radial orders are not considered in the fitting. Nor we have considered the last two radial orders, most likely affected by aliasing. Despite the good fitting achieved in both cases shown in Figure 2, we should note that the seeing estimation for the case of binning #3 and 66 controlled modes is actually under-estimated. Indeed, the mismatch between the PSF size during the calibration of the interaction matrices (diffraction-limited) and the actual PSF size in closed-loop operation at WFS wavelengths is responsible for the underestimation of the modal residuals, which in general leads to an underestimation of the seeing value. We have found from our data analysis that this underestimation is non-negligible for datasets taken with star magnitudes $M_R >13$. We are currently working on an improvement of the on-line seeing estimation method in which the modal residuals are iteratively estimated.

## 4. ON-SKY COMMISSIONING RESULTS

We will now present the overall performance analysis of the FLAO#1 system based on the data collected during the commissioning nights. The SR was estimated from the long-exposure PSFs measured with the InfraRed Test Camera[9] (IRTC) using an H-band filter with an effective central wavelength of 1.60μm. Total integration times varied from ~30s for high-flux PSF measurements up to ~120s for the low-flux ones. The camera was operated in the narrow Field-of-View (FoV) mode with a pixel scale of 10 mas/pixel.

Figure 3 summarizes the performance results attained by the FLAO#1 system. A total of 597 Strehl Ratio estimations at star magnitudes from ~7.5 to ~18 are shown in the plot. Seeing values estimated from AO real-time data are coded in color. Points ($M_R>13$) for which seeing values are strongly under-estimated with this method are shown in black. Figure 3 also indicates which system configuration (i.e. binning mode) has been used for each acquisition. Finally, the expected performance estimated from numerical simulations and for different seeing values (0.6", 0.8", 1.0", 1.2", and 1.5") is also shown in the plot. It is important to note that the measured SR values are in accordance with the simulated ones, showing that the FLAO#1 system meets the expected performance.

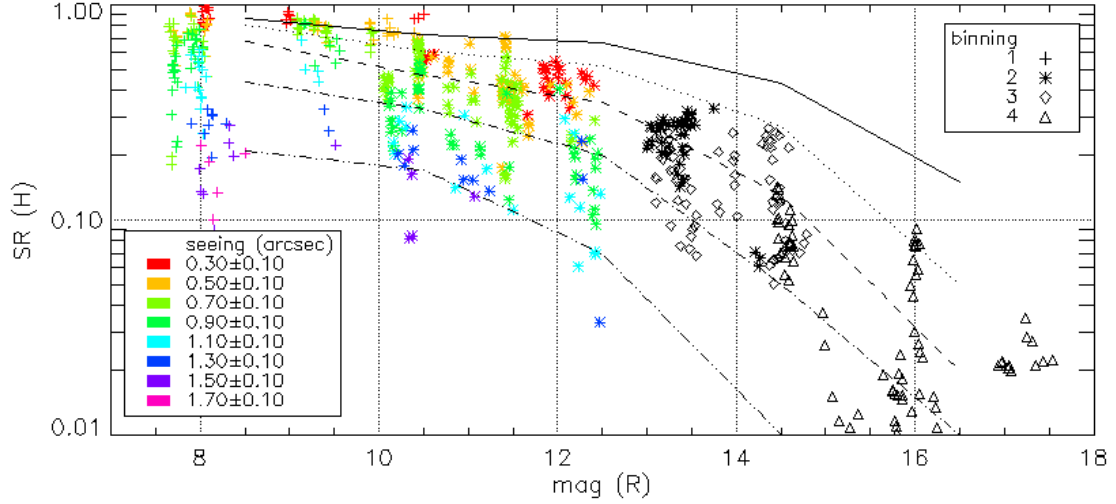

Figure 3. (Color online) Summary plot of FLAO#1 performance commissioning results. Strehl Ratio in H band versus star magnitude (R) and for different seeing values (estimated from AO real-time data). Points ($M_R$ >13) for which seeing values are strongly under-estimated are shown in black. Different binning modes are indicated with symbols. Lines correspond to expected performance from numerical simulations with different seeing values; from top to bottom: 0.6", 0.8", 1.0", 1.2", and 1.5".

Figure 4 shows some examples of AO-corrected PSFs sampling the range of star magnitudes ($8<M_R<17$), and for a seeing value of about 0.8arcsec. Note that a diffraction-limited resolution of 40mas is achieved down to magnitude ~12. In the next subsections we will discuss some particular cases.

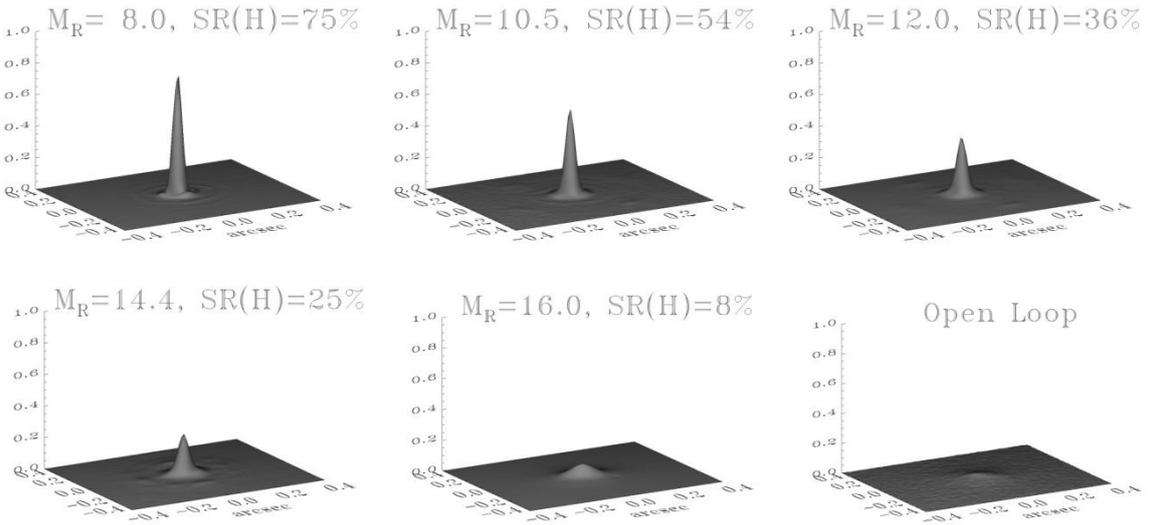

Figure 4. Examples of AO-corrected PSFs (normalized to the diffraction-limited peak) acquired at different star magnitudes. The system configurations are listed in Table 1. The seeing value for all of these acquisitions was about 0.8 arcsec. An OL PSF is also shown for comparison.

## 4.1 Performance at the bright end: high-contrast imaging

Figure 5 shows an example of a high-order corrected PSF in H band with a SR of >80%. The equivalent R magnitude of the GS (HD175658) is 6.5 and the on-line estimated seeing oscillated between 0.6 and 0.8 arcsec. The AO loop was running at 1 kHz controlling 400 KL modes. Some residual tip/tilt vibrations at ~13.4Hz of 6 to 8 mas rms were present. Figure 5 also shows the radial-averaged profiles of the AO-corrected and the diffraction-limited PSFs. It is important to note that a contrast better than $10^4$ is achieved at a radial distance of 0.4 arcsec radius from the central peak, just before the turbulence residual halo occurring at ~0.47 arcsec. This image confirms on sky for the first time the deep annular region where the non-aliased correction offered by the pyramid sensor achieves the maximum contrast[10].

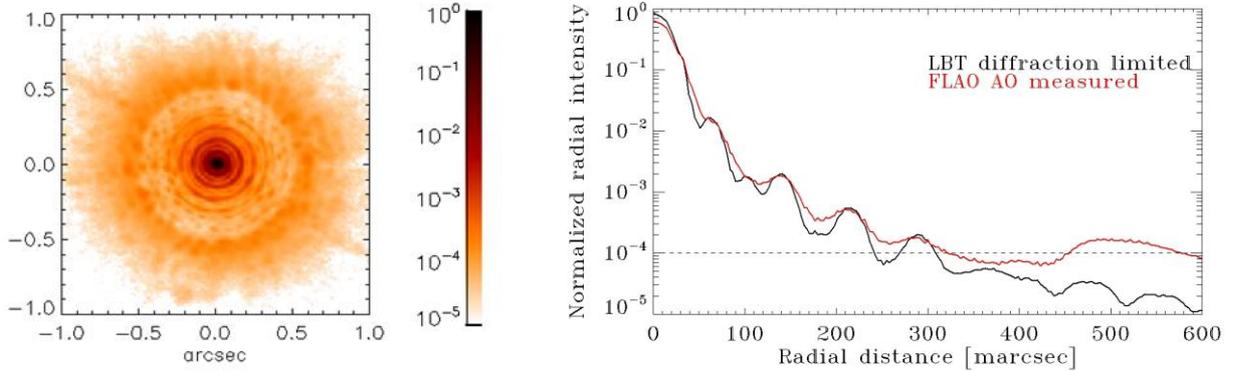

Figure 5. (*Left*) High-order AO-corrected PSF using a bright star ($M_R$ = 6.5) under median seeing conditions (seeing~0.7"±0.1"). (*Right*) Comparison between the diffraction-limited PSF (black full curve) and the AO-corrected PSF profiles (red full curve). Profiles are normalized to the diffraction-limited peak.

## 4.2 Performance at the faint end

Let us now discuss the performance of the FLAO system with faint reference stars, close to the current limiting magnitude. Figure 6 shows an example of the results obtained at $M_R$= 17.2. The star observed was LP154-66 ($M_R$=13.20, $M_H$=10.17), and a neutral density filter was placed on the WFS channel to emulate a 17.2 R-magnitude star. The system was operated with binning mode #4 and controlling only 10 modes giving a SR(H) of ~3.5%. The FWHM of the AO-corrected PSF is reduced a factor of 1.7 with respect to the seeing-limited one, and their peaks ratio equals 2.6.

A seeing estimate for this acquisition was not available from the DIMM. We can nevertheless use the FWHM of the seeing-limited PSF and, assuming a given outer scale value, estimate the seeing as in Tokovinin(2002)[11]. Following this method, and considering an outer scale of $L_0$ = 40m, we get a seeing estimate of 0.72" (@ 500nm) for this observation.

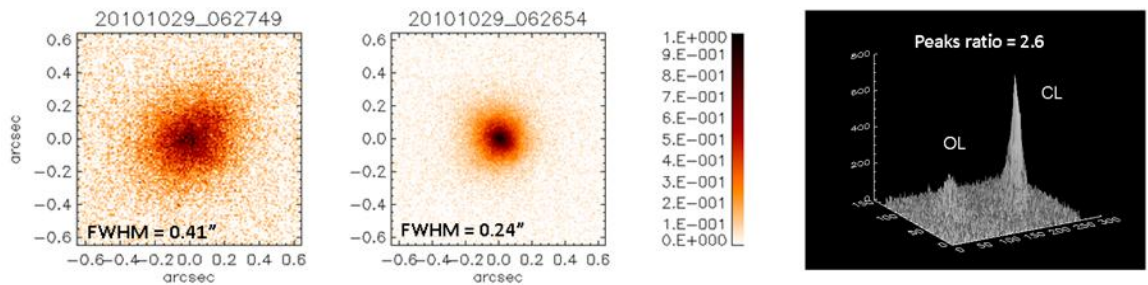

Figure 6. Example of the FLAO performance in H band at its faint end ($M_R$= 17.2). (*Left*) Seeing-limited image. (*Middle*) AO-corrected image. System parameters: binning mode #4, 10 controlled modes, 100Hz, and a modulation of $\pm 6\lambda_s/D$. Both images are normalized to their peak intensities. (*Right*) Images placed side by side to show the gain in energy concentration. These images were produced from 5 IRTC frames of 3s exposure time each.

## 5. NGS AND LGS ADAPTIVE OPTICS SYSTEMS AFTER LBT RESULTS

We discuss in this section the case of adaptive correction on a small FoV, more precisely the FoV limited by the size of the isoplanatic patch of the atmosphere. Recall that the development of Laser Guide Star (LGS) systems to create suitable reference stars for adaptive optics was mostly driven by the relative inefficiency of the first NGS-based AO systems for science applications. Indeed, the efficiency in exploiting the star flux was in the past lower, with the consequence of achieving a limited sky coverage of the order of one to a few percent, in most cases. Also, the fast read-out CCDs were noisier and the efficiency at low flux levels of the wave-front sensors had to be improved.

In the case of the LBT, the quest for efficiency led to the development of its two AO key components. On one hand, the ASM allows to minimize the number of reflections giving the WFS the possibility to catch the light after only three reflections (primary mirror, ASM, and the folding tertiary); on the other hand, the modulated pyramid allows for a gain of the order of 1-2 limiting magnitudes with respect to commonly used Shack-Hartmann sensors[12][13]. In terms of system efficiency, a future upgrade consists in replacing the current WFS camera (E2V CCD 39) with a Low Light Level CCD (LLLCCD) also called Electron Multiplied CCD (EMCCD) as proposed by Carbillet and Riccardi (2010)[14].

### 5.1 Improving NGS AO performance and Sky Coverage with an LLLCCD

We have evaluated the performance of an LBT-like system using an LLLCCD with the numerical simulator we developed for FLAO performance verification[7]. The main parameters of the LLLCCCD we simulated are the following: 0.1 e$^-$ RMS noise from Clock Induced Charges (CIC), 0.03 e- RMS of RON at 1kfps. The main atmospheric parameters used in the simulations are: seeing of 0.6 arcsec, outer scale of $L_0$=40m, and a wind speed of 15m/s.

Results of the simulations in terms of SR in K band (2.2μm) as a function of the GS magnitude are reported in Figure 7 for the FLAO system with a CCD39 and with an LLLCCD. As a comparison, we report on the same plot a typical performance curve for the LGS AO system of the Keck telescope, one of the most successful LGS-based AO systems[15]. Performance curves are reported for a seeing value of 0.6 arcsec (median seeing in Mauna Kea[16]). We recall here briefly that the LGS system needs to use an NGS as the reference for the tip/tilt signal that is not provided by the LGS[17]. However, in this case the NGS is used to measure only the overall tip/tilt and so it can be fainter than the usual reference star for an NGS system. Figure 7 shows that the LGS system has a better limiting magnitude with respect to the FLAO NGS cases. For instance, for a SR of 10% in K band, we obtain 18.4 and 19.0 limiting magnitudes for NGS+LLLCCD and LGS respectively. On the other hand, the NGS+LLLCCD system performance is higher than the LGS system up to magnitude 16.8 reaching a SR higher than 70% up to magnitude 15.2. We note that one of the main limitations of LGS performance in the bright end is focus anisoplanatism error[18].

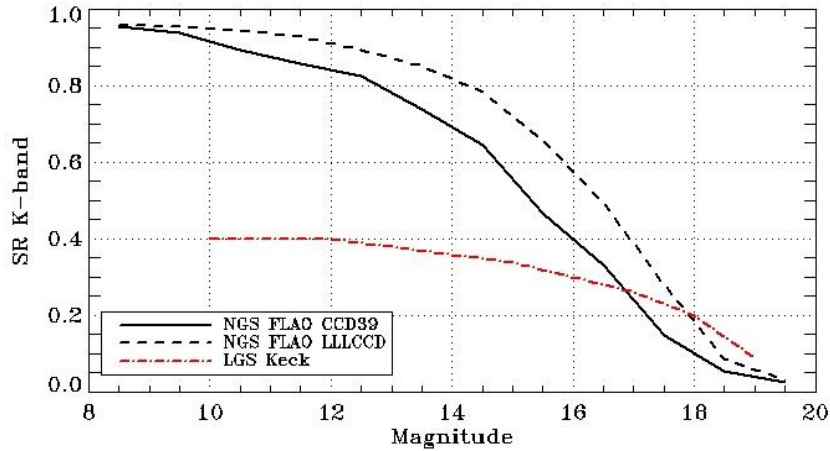

Figure 7. Simulated performance (seeing of 0.6 arcsec) in terms of SR in K band versus star magnitude for the LBT FLAO system and of the improved version using an LLLCCD. The performance of the Keck LGS AO system is shown for comparison.

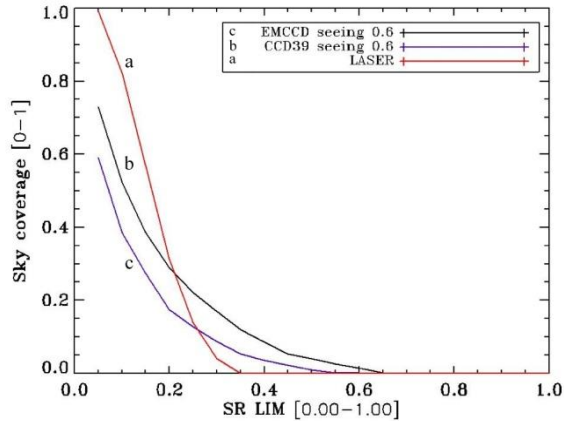

Figure 8 Sky coverage results for the Steidel & al (2003) sample of galaxies using the actual LBT FLAO system (curve c), the improved version using LLLCCD (curve b) and the Keck AO LGS system (curve a). The analyses were made under 0.6"seeing values.

We computed the Sky Coverage (SC) for the two systems in question (NGS+LLLCCD and LGS systems) for a seeing value of 0.6 arcsec with isoplanatic and isokinetic angles of 40 and 73 arcsec respectively in K band. The faint limiting magnitudes considered imply very high sky coverage at low galactic latitudes for both NGS and LGS systems and the comparison is not so interesting. We decided to compute the SC for an astronomical sample of targets located at high galactic latitude. In particular, we selected a sample of about 850 galaxies at z=3 from the Steidel catalog[19] and computed the SR performance according to the plot already shown in Figure 7. The results are shown in Figure 8 where the SC achieved for a given SR is reported for both systems. As a reference, the same curve for the NGS+CCD39 system is also shown in the plot. The result here is that for a required SR of 10% in K band the LGS system achieves an 84% SC while the NGS+LLLCCD system a 54% SC.

On the other hand, for a required SR higher than 25%, the NGS system has a better SC than the LGS system, being on average about 15%. In summary, the above computations show that an LBT-like AO system upgraded from the FLAO system being already demonstrated on sky, has comparable performance to an LGS system as SC is concerned, and has definitely better performance when higher SR values are required.

### 5.2 Future developments for AO systems

An interesting point about NGS and LGS comparison is the use of adaptive optics systems at wavelengths shorter than the K and H bands like J or even visible wavelengths. Because of the focus anisoplanatism effect, the single LGS correction becomes less and less effective when going to shorter wavelengths[18]. Hence, such a single laser system for an 8m class telescope is limited to correction in K and H band. This limitation does not exist for the NGS system that can be pushed to shorter wavelengths provided a sufficient number of photons and actuators exist. The LBT telescope is a first example of this, having achieved about 50% SR in J band and 20% of SR at 850nm. The scaling of LBT on-sky results to shorter wavelengths is shown in Figure 9. Let us mention that the 6.5 Magellan telescope[20] is in the testing phase of an LBT-like system aimed at visible observations. A similar system could be developed for the ESO UT4 telescope being equipped in the next future with an adaptive secondary having more than 1100 actuators[21]. Such a system could deliver 15 mas FWHM at 0.6 μm. In the LGS case, visible AO requires to solve the focus anisoplanatism issue using more than one LGS and so in turn using more than one NGS for the measurements of wavefront low-order modes[22]. On the other hand, NGS systems do no require major changes to be operational at shorter wavelengths as already discussed.

Finally, if we move to AO systems for ELTs, the design of an LGS system becomes very complex considering the number of LGS and NGS needed, the needs to deal with the spot elongations effect, and several other issues arising from the use of LGS such as the variability of the sodium density profile, etc. On the contrary, an NGS-based system for an ELT is built exactly as a system for an 8m telescope, the only difference being the number of actuators and subapertures. Such considerations summed to the easiness of operation and reduced cost of an NGS-based AO system suggests rethinking the use of LGS in AO to improve system SC and reliability for 8m and ELT class telescopes. This at least for AO systems having the corrected FoV limited by the atmospheric isoplanatic patch.

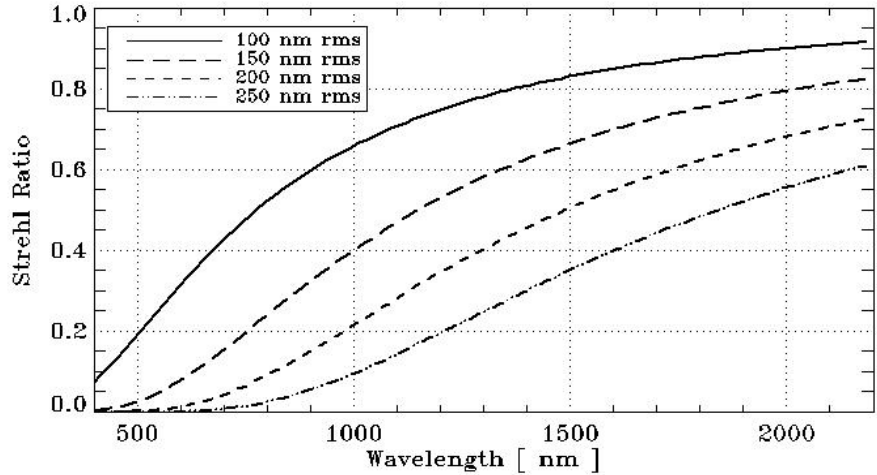

Figure 9. Strehl Ratio scaling as a function of the wavelength. Each curve represents a different level of wave-front correction corresponding to SR values in H band of 85%, 70%, 55%, and 40%.

## 6. CONCLUSIONS

The First Light AO system (FLAO#1) of the 2x8.4m LBT telescope has been commissioned on sky in the period from June 2010 to August 2011. The achieved results —such as H band SR higher than 80%, image contrast better than $10^4$, and loop closure down to 17.5 magnitude star— sets a new standard for ground based astronomical AO systems. At the same time these results show that adaptive secondary mirrors and pyramid sensors are mature technologies to be used in the design of future AO systems for 8m and ELT class telescopes.

Finally, an elaboration of the LBT experience does show that an LBT-like system using an LLLCCD can achieve, in the case studied of extragalactic objects, a sky coverage of 54% comparable to the one obtained with an LGS of 84%. Moreover, a better sky coverage at SR higher than 25% can be achieved by the NGS system. These results, coupled with relevant limitations of LGS systems like cone effect and the need for low-order modes sensing with an NGS, suggest that the choice of using LGS systems instead of NGS systems to improve AO system efficiency should be revisited.

## ACKNOWLEDGEMENTS

The authors would like to thank the LBTO personnel for their continuous and timely support during the integration and commissioning of the system.